\documentclass [11pt,letterpaper]{JHEP3}
\usepackage{graphicx}
\usepackage{epsfig}
\usepackage{cite}	
\usepackage{bm}		
\advance\parskip 1.5pt plus 1.0pt minus 1.0pt	
\def\coeff#1#2{{\textstyle {\frac {#1}{#2}}}}
\def\half{\coeff 12}

\def\tr{{\rm tr}}

\def\Nc{N_{\rm c}}

\def\Z{{\mathbb Z}}

\def\Dslash{{\rlap{\raise 1pt \hbox{$\>/$}}D}}

\def\parent{{\rm (p)}}
\def\daughter{{\rm (d)}}

\preprint {UW/PT 05--12\\BUHEP 05--09\\NSF-KITP--05--28}
\title
    {%
    Can large $\bm \Nc$ equivalence between
    \boldmath supersymmetric Yang-Mills theory
    and its orbifold projections be valid?
    }%
\author
    {%
    Pavel Kovtun\footnotemark,
    Mithat \"Unsal\footnotemark,
    and Laurence G.~Yaffe\footnotemark
    \\${}^*$KITP, University of California, Santa Barbara, CA 93106
    \\${}^\dagger$Department of Physics, Boston University, Boston, MA 02215
    \\${}^\ddagger$Department of Physics, University of Washington,
    Seattle, Washington 98195--1560
    \\Email:
    \parbox[t]{2in}{\email {kovtun@kitp.ucsb.edu},\\
		    \email {unsal@buphy.bu.edu},\\
		    \email {yaffe@phys.washington.edu}}
    }%

\abstract
    {%
    In previous work, we found that
    necessary and sufficient conditions for large $\Nc$ equivalence between
    parent and daughter theories,
    for a wide class of orbifold projections of $U(\Nc)$ gauge theories,
    are just the natural requirements that
    the discrete symmetry used to define the projection
    not be spontaneously broken in the parent theory,
    and the discrete symmetry permuting equivalent gauge group factors
    not be spontaneously broken in the daughter theory.
    In this paper, we discuss the application of this result to
    $\Z_k$ projections of ${\cal N}\,{=}\,1$ supersymmetric
    Yang-Mills theory in four dimensions,
    as well as various multi-flavor generalizations.
    $\Z_k$ projections with $k > 2$ yielding chiral
    gauge theories violate the symmetry realization conditions needed
    for large $\Nc$ equivalence, due to the spontaneous symmetry breaking
    of discrete chiral symmetry in the parent super-Yang-Mills theory.
    But for $\Z_2$ projections,
    we show that previous assertions of large $\Nc$ inequivalence,
    in infinite volume, between the parent and daughter theories
    were based on incorrect mappings of vacuum energies,
    theta angles, or connected correlators between the two theories.
    With the correct identifications,
    there is no sign of any inconsistency.
    A subtle but essential feature of the connection between
    parent and daughter theories involves multi-valuedness
    in the mapping of theta parameters from parent to daughter.
    }%


\begin {document}

\setlength{\baselineskip}{1.10\baselineskip}

\section {Introduction}

Orbifold projection is a technique for constructing ``daughter''
theories starting from some ``parent'' theory, by retaining only
those fields which are invariant under a chosen discrete symmetry group
of the parent theory.
In some cases involving orbifold projections of $U(\Nc)$ gauge theories,
the large $\Nc$ limits of parent and daughter theories may coincide
\cite {Bershadsky-Johansen,Schmaltz,Strassler,Dijkgraaf-Neitzke-Vafa,KUY1}.
More precisely, for a wide class of projections yielding daughter
theories with $U(N)^k$ product gauge groups, it has recently been shown,
rigorously, that the leading large-$\Nc$ dynamics in the sector of
the parent theory invariant under the discrete symmetry used to define the
projection coincides with the large $\Nc$ dynamics in the sector of the
daughter theory which is invariant under permutations of
equivalent gauge group factors \cite {KUY2}.
The ground states of each theory will lie in these symmetry sectors
(which we will refer to as {\em neutral}\/%
\footnote
    {
    In the parent theory, neutral operators (or states)
    are gauge invariant operators (or states) which are
    invariant under the chosen projection symmetry.
    In the daughter theory, neutral operators/states are
    gauge invariant operators/states which are also invariant
    under any additional global symmetries, such as permutations of different
    gauge group factors, which are remnants of gauge symmetries
    in the parent theory.
    In string theory literature,
    non-neutral operators of the daughter theory
    are referred to as ``twisted''.
    })
provided the symmetries defining the projection
are not spontaneously broken in the parent theory, and the
gauge group permutation symmetries are not spontaneously broken
in the daughter.
These symmetry realizations constraints are both
necessary and sufficient for the validity of a non-perturbative
equivalence between parent and daughter theories relating the
leading large $\Nc$ behavior of physical properties such as
mass spectra and scattering amplitudes of neutral particles
\cite {KUY2}.

In this paper, we examine the application
of this result to $\Z_k$ projections of ${\cal N}\,{=}\,1$
supersymmetric Yang-Mills theory in four dimensions
with gauge group $U(2N)$,
or its multi-flavor generalizations.
Much of our discussion will focus on the simplest case of
a $\Z_2$ projection of ${\cal N}\,{=}\,1$ super-Yang-Mills,
which yields a non-supersymmetric $U(N)\times U(N)$ gauge theory.
This example has been discussed previously by a number of authors
\cite{Strassler,Gorsky-Shifman,Tong,ASV}.
Gorsky and Shifman \cite {Gorsky-Shifman},
and Armoni, Shifman, and Veneziano \cite {ASV}
argued that non-perturbative large $N$ equivalence fails in this case,
based on apparent mis-matches in topological susceptibility,
and of instanton zero modes when the theory is compactified
on $T^4$ or $R^3 \times S^1$.
Tong \cite{Tong} also considered this example
when compactified on $R^3\times S^1$
and demonstrated, explicitly, that the orbifold equivalence fails
when the compactification radius is sufficiently small.
Based on this and closely related examples, Ref.~\cite{ASV}
suggests that large-$N$ orbifold equivalence may generally fail.

We will show that the apparent discrepancies
discussed in Refs.~\cite{Gorsky-Shifman,ASV}
between these parent and daughter theories, formulated on $R^4$,
are illusory, and result from incorrect mapping of observables
between the two theories.
In particular, in these papers it was assumed that a valid equivalence
would imply that the vacuum energy densities
and topological susceptibilities of parent and daughter theories
(when supersymmetry is softly broken)
would coincide.
This is incorrect.
As we discuss,
in this example the vacuum energy density of the parent theory
must be compared with twice the energy density of the daughter,
while the topological susceptibility of the parent should be half
that of the daughter (because the $\theta$ angle of the parent
must be identified with twice the $\theta$ angle of the daughter theory%
\footnote
    {
    This is the same relation which results from the
    mapping of the holomorphic coupling $\tau \equiv 4\pi/g^2 + i\theta/2\pi$
    in examples where both parent and daughter are supersymmetric
    \cite{Erlich-Naqvi}.
    For a $\Z_2$ projection, coinciding 't Hooft couplings requires
    $\tau_{\rm parent} = 2 \tau_{\rm daughter}$,
    implying that $\theta_{\rm parent} = 2 \theta_{\rm daughter}$.
    \label{fn:holomorphic}
    }).

With the correct mapping between parent and daughter theories,
all evidence for any inconsistency of the large-$N$ orbifold equivalence,
in the uncompactified theories, disappears.
A subtle feature, which is essential for understanding the correspondence
between individual vacua, or domain walls, in the two theories, involves
multi-valuedness in the mapping of theta parameters between parent
and daughter theories.
This is discussed in some detail.
We argue that there is no sign of any inconsistency with large $N$ equivalence
between these two theories, when formulated on $R^4$,
and no evidence suggesting spontaneous symmetry breaking of the
$\Z_2$ symmetry interchanging the $U(N)$ factors in the daughter theory.
However, as Tong \cite {Tong} showed quite explicitly,
if the theories are compactified on $R^3\times S^1$
then at sufficiently small radius the symmetry interchanging
the $U(N)$ factors in the daughter theory does break spontaneously,
thereby violating the necessary conditions for large $N$ orbifold
equivalence.

We also examine extensions of these theories involving
additional adjoint fermions, and explain why a mis-match
in the total number of Goldstone bosons,
noted in Ref.~\cite{Gorsky-Shifman},
is also not evidence for failure of large-$N$ orbifold equivalence
in these examples.
We find that the number of Goldstone bosons in the relevant neutral
symmetry channels, to which the equivalence applies, do match.

We briefly discuss $\Z_k$ projections with $k > 2$,
which can yield $U(N)^k$ chiral gauge theories with bifundamental fermions.
Large $N$ equivalence between the parent and daughter theories
does fail for these projections, due to the spontaneous symmetry breaking
of discrete chiral symmetry in the parent theory and in complete accord
with the symmetry realization conditions discussed in Ref.~\cite{KUY2}.
We also comment on the case of $\Z_2$ projection of
${\cal N} \,{=}\, 4$ supersymmetric Yang-Mills theory yielding
a non-supersymmetric $U(N)^2$ gauge theory with adjoint scalars
and bifundamental fermions.
This case, which has received considerable previous attention
(see, for example,
Refs.~\cite{Tseytlin:1999ii,Adams:2001jb,Klebanov:1999ch,Nekrasov:1999mn,
Klebanov:1999um,Dymarsky:2005uh})
appears to be an example of large $N$ inequivalence between parent and daughter
theories due to spontaneous $\Z_2$ symmetry breaking in the daughter.
However, this does not imply that any analogous symmetry breaking
must occur in the ${\cal N} \,{=}\, 1$ case.

As this paper was (finally) nearing completion,
the recent preprint \cite {AGS} appeared.
This work also discusses the consistency of large $N$ orbifold equivalence
for the $\Z_2$ projection of ${\cal N}\,{=}\,1$ super-Yang-Mills,
but reaches radically different conclusions from ours,
namely manifest inconsistency of large $N$ equivalence
based on multiple different lines of reasoning.
In an addendum to the conclusion of this paper,
we briefly examine the assertions of Ref.~\cite {AGS}
and find that the multiple apparent inconsistencies
discussed in that work
can all be traced to incorrect mapping of operators
and correlation functions between the two theories.
Once the correct mappings are used
(as spelled out below and in Ref.~\cite {KUY2}),
all evidence for failure of large $N$ equivalence in this example
(and hence all evidence for spontaneous breaking of the $\Z_2$ symmetry
in the daughter theory) disappears.


\section {\boldmath $\Z_2$ projection of ${\cal N}\,{=}\,1$
	supersymmetric Yang-Mills theory}
\label {sec:SYM}

As has been recognized previously
\cite{Strassler,Gorsky-Shifman,ASV,Tong},
an interesting example of orbifold
projection is provided by the $\Z_2$ projection
of ${\cal N}\,{=}\,1$ supersymmetric Yang-Mills theory in
four dimensions.
The parent theory is a $U(2N)$ gauge theory with a massless
Weyl fermion in the adjoint representation.%
\footnote
    {%
    The difference between $SU(2N)$ and $U(2N)$ gauge groups
    is irrelevant in the large $N$ limit.
    }
The daughter theory is a $U(N)\times U(N)$ gauge theory with
a bifundamental Dirac fermion.
Both parent and daughter theories are asymptotically free,
confining gauge theories with massless fermion fields.
The parent theory is supersymmetric, but the daughter theory is not.
Therefore the equivalence, if true, would dictate exact relations
in the large $N$ limit
between mass spectra, correlators, and scattering amplitudes of a
supersymmetric parent theory and its non-supersymmetric daughter
\cite{Strassler}.%
\footnote
    {
    For example, unbroken supersymmetry in the parent theory implies
    that massive scalar particles in the smallest supermultiplet
    must be degenerate in mass with both a pseudoscalar and a spin 1/2 fermion.
    Only bosonic particles lie in the neutral sector of the parent theory.
    Hence, the mass spectrum of the daughter theory, in the
    $N \to\infty$ limit, must also exhibit parity doubled scalars
    and pseudoscalars,
    provided the $\Z_2$ symmetry interchanging gauge group factors
    is not spontaneously broken.
    }

The authors of Refs.~\cite{Gorsky-Shifman, ASV} argued that the
topological susceptibilities of the two theories,
in infinite volume,
do not match, thus invalidating the equivalence.
More precisely, they considered the theories with a small fermion mass
turned on, and found that if the vacuum energies
of the two theories coincide at $\theta = 0$,
then the curvatures of the vacuum energies with respect to $\theta$
do not match.
We wish to reexamine this comparison between parent
and daughter theories.
The action density of the parent theory may be written as
\begin{equation}
    {\cal L}^\parent
    =
    -\frac{1}{2g_p^2} \, \tr \, F_{\mu\nu} F^{\mu\nu}
    +\frac{\theta_p}{16\pi^2} \, \tr\,F_{\mu\nu}\widetilde{F}^{\mu\nu}
    + \frac 1{g_p^2} \, \tr
	\left[
	    \bar\lambda_{\dot\alpha} \,
	    (\bar\sigma^\mu)^{\dot\alpha\alpha} D_\mu
	    \lambda_\alpha
	    + \half m_p \, (\lambda^\alpha \lambda_\alpha {+}
			\bar\lambda_{\dot\alpha} \bar\lambda^{\dot\alpha})
	\right] .
\end{equation}
The traces are over $2N\times 2N$ matrices
with $F_{\mu\nu} \equiv F_{\mu\nu}^a \, t^a/(2i)$,
and $\lambda_\alpha \equiv \lambda_\alpha^a \, t^a/(2i)$
an adjoint representation Weyl fermion.
With an anomalous chiral rotation,
the theta parameter may be moved into the fermion mass term.
In the massless limit,
the non-anomalous discrete $\Z_{4N}$ chiral symmetry of the parent
theory is spontaneously broken to its $\Z_2$ subgroup by the
vacuum fermion condensate, giving rise to $2N$ distinct vacua
which are related to each other by shifts in $\theta_p$ by
multiples of $2\pi$.
The unbroken $\Z_2$ symmetry is just $\lambda \to -\lambda$,
corresponding to conservation of fermion number modulo two.
When a small mass is added,
the resulting vacuum energy density is
\begin{equation}
  {\cal E}^\parent
  =
  \frac{m_p}{2 g_p^2} \left[
     \langle \tr\,\lambda\lambda\rangle \, e^{i\theta_p/(2N)}
    +\langle \tr\,\bar\lambda\bar\lambda\rangle \, e^{-i\theta_p/(2N)}
    \right]
  + O(m_p^2) \,,
\label{eq:dE-parent}
\end{equation}
where the phase of $\langle \tr\,\lambda\lambda\rangle$
is an integer multiple of $\pi/N$;
the particular value depends on which of the $2N$
vacua of the massless theory is under consideration.
Changing $\theta_p \to \theta_p + 2\pi$
is the same as moving from one vacuum state to the next.
The magnitude of $\langle \tr \, \lambda\lambda \rangle$
is $O(N \Lambda^3)$, where $\Lambda$ is the conventional
non-perturbative scale of the theory \cite {Schmaltz,Witten}.
Therefore,
the energy density (\ref {eq:dE-parent}) is $O(N^2)$
[since $g_p^2$ is $O(1/N)$ at fixed 't Hooft coupling]
as required for the leading large $N$ behavior in a
$U(2N)$ gauge theory.

The $\Z_2$ orbifold projection eliminates all degrees of
freedom except those invariant under the combination
of a global gauge transformation by $\gamma \equiv {\rm diag}(1_N,-1_N)$
(with $1_N$ denoting an $N \times N$ identity matrix),
combined with a fermion sign flip, $\lambda \to -\lambda$.
This forces the gauge field $A_\mu$ to be block-diagonal,
reducing it from a $U(2N)$ gauge field to two independent $U(N)$
gauge fields,
whose field strengths we will denote as $F_{\mu\nu}^1$ and $F_{\mu\nu}^2$.
The extra sign change for the fermion means that
$\lambda$ becomes block off-diagonal,
so that the daughter theory contains two Weyl fermions transforming
as bifundamentals under $U(N)\times U(N)$,
which we will denote as $\lambda^1$ and $\lambda^2$.
(See Refs.~\cite{Gorsky-Shifman,ASV} for a more detailed explanation.)
Under this projection,
\begin {eqnarray}
    \tr \, F F
    &\to&
    \tr \, (F^1 F^1 + F^2 F^2) \,,
\\
    \tr \, \lambda\lambda
    &\to&
    \tr \, (\lambda^1 \lambda^2 + \lambda^2 \lambda^1)
    =
    2 \, \tr \, \lambda^1 \lambda^2  \,,
\label {eq:map lamlam}
\end {eqnarray}
where traces on the left are over $2N \times 2N$ matrices,
while those on the right are $N \times N$.
The action density of the daughter theory has the form
\begin{equation}
  {\cal L}^\daughter
  =
  \sum_{j=1}^{2}
  \Bigl\{
      -\frac{1}{2g_d^2} \, \tr \, F^j F^j
      +\frac{\theta_d}{16\pi^2} \, \tr\, F^j\widetilde{F}^j
      + \frac 1{g_d^2} \, \tr \,
	    \bar\lambda^j \, \bar\sigma^\mu D_\mu \lambda^j
   \Bigr\}
    + \frac {m_d}{g_d^2} \, \tr \,
    (\lambda^1 \lambda^2 {+} \bar\lambda^2 \bar\lambda^1) \,,
\end{equation}
where the fermion covariant derivative includes the appropriate
coupling to both $U(N)$ gauge fields.
This theory has a non-anomalous discrete $\Z_{2N}$ chiral symmetry,
which is also expected to break spontaneously due to the formation
of a fermion bilinear condensate, leaving an unbroken
$\Z_2$ subgroup (corresponding to fermion number modulo two).
In addition, the daughter theory has a $\Z_2$ ``theory space''
symmetry which interchanges the two $U(N)$ gauge groups
({\em i.e.}, $F^1 \leftrightarrow F^2$,
$\lambda^1 \leftrightarrow \lambda^2$).
One may again rotate the theta parameter into
the fermion mass term.
The resulting shift in the vacuum energy density is
\begin{equation}
  {\cal E}^\daughter
  =
    \frac{m_d}{g_d^2}
    \left[
        \langle \tr\,\lambda^1 \lambda^2 \rangle \, e^{i\theta_d/N}
       +\langle \tr\, \bar\lambda^2 \bar\lambda^1 \rangle \,
       e^{-i\theta_d/N}
    \right]
  + O(m_d^2) \,,
\label{eq:dE-daughter}
\end{equation}
where the phase of $\langle \tr\,\lambda^1\lambda^2\rangle$
is now an integer multiple of $2\pi/N$,
with the particular value depending on which of the $N$ vacua of the
massless daughter theory is under consideration.
Changing $\theta_d \to \theta_d + 2\pi$
is again the same as moving from one vacuum state to the next.

\subsection {Parameter mapping between parent and daughter}

The parameters of the daughter theory are {\em not} determined by
equating the daughter theory action with the parent theory action,
after replacing fields with their projected forms.
Rather, as shown in Refs.~\cite{KUY1,KUY2},
one must equate the parent action, after projecting fields,
with {\em twice} the daughter action,
\begin {equation}
    {\cal L}^\parent \to 2 \, {\cal L}^\daughter \,.
\label {eq:hphd}
\end {equation}
(More generally, for a $\Z_k$ orbifold projection the parent action,
after projecting fields, must be identified with $k$ times the
daughter action.)
This implies that the correct relations between parameters of the
parent and daughter theories are:
\begin{eqnarray}
  g_p^2 &=& \half \, g_d^2 \,,\qquad
\label {eq:map coupling}
  \theta_p = 2 \, \theta_d \,,\qquad
\label {eq:map theta}
  m_p = m_d \,.
\label {eq:map m}
\end{eqnarray}
Both the gauge coupling constant
and the vacuum angle of the daughter theory
differ from their parent theory counterparts.
The relation (\ref {eq:map coupling}) for the gauge couplings
is equivalent to equality of 't Hooft couplings in parent
and daughter theories;
the necessity of this condition for large $N$ orbifold equivalence
may be seen in perturbation theory \cite{Bershadsky-Johansen}.
Equality of fermion masses is obvious, and may also be seen
in perturbation theory.
The relation (\ref {eq:map theta}) between parent and daughter theta angles
cannot be seen in perturbation theory but,
as noted in footnote \ref{fn:holomorphic}, this relation
also follows from the mapping of the holomorphic coupling in cases
where both the parent and daughter theories are supersymmetric.

\subsection {Energy densities and topological susceptibilities}

The relation (\ref{eq:hphd}) connecting parent and daughter
theory actions also applies to their Hamiltonians \cite {KUY2}.
Non-perturbative large $N$ equivalence between the parent and
daughter theories will be valid provided the $\Z_2$ theory space
symmetry is not spontaneously broken in the daughter.%
\footnote
    {
    We assume that the $\lambda\to -\lambda$ [or $(-1)^F$]
    symmetry in the parent, used to define the orbifold projection,
    cannot be spontaneously broken.
    This is automatic provided Lorentz invariance is unbroken.
    Since gauge symmetries can never be spontaneously broken \cite{Elitzur},
    it is only $\Z_2$ symmetry breaking in the daughter which
    could prevent large $N$ equivalence between parent and daughter.
    }
In this case, their vacuum energies must satisfy the same relation
at large $N$,
\begin {equation}
    {\cal E}^\parent = 2 \, {\cal E}^\daughter  \times [1 + O(1/N^2)] \,.
\label {eq:map E}
\end {equation}
The factor of two in this relation should not be surprising.
One way to understand it physically is to consider the
free energy at temperature $T$, instead of the vacuum energy.
At asymptotically high temperatures, the free energy density
effectively counts the number of degrees of freedom,
half of which are removed by the orbifold projection.%
\footnote
    {
    Explicitly, the free energy density in the parent is
    $
	{\cal F}^\parent
	=
	-{\pi^2 \over 24} \, (2N)^2 \> T^4 \> [1 + O(2N g_p^2)]
    $,
    while in the daughter
    $
	{\cal F}^\daughter
	=
	-{\pi^2 \over 24} \, (2N^2) \> T^4 \> [1 + O(N g_d^2)]
    $.
    Hence ${\cal F}^\parent = 2 \, {\cal F}^\daughter$
    at high temperature.
    }

Using the relations (\ref {eq:map coupling})
for the coupling constants,
plus (\ref {eq:map lamlam}) for the fermion bilinear,
one may immediately see that the expressions
(\ref{eq:dE-parent}) and (\ref{eq:dE-daughter})
for the first order vacuum energy shift due to a fermion mass
are completely consistent with the relation (\ref {eq:map E})
between vacuum energy densities.%
\footnote
    {
    If fermion mass is sufficiently large, then both parent and
    daughter theories have unique ground states,
    and large $N$ parent-daughter equivalence is guaranteed to hold
    throughout the large-mass phase.
    The only way the equivalence could fail in the massless limit
    is if the daughter theory has a $\Z_2$ symmetry breaking phase transition
    at some non-zero mass $m_*$.
    This critical mass would have to be of order of the strong scale
    of the theory, $m_* = C \, \Lambda$.
    But as we discuss in the remainder of this section,
    available information regarding the dynamics of the daughter theory
    is consistent with the expectation that $C=0$
    and the absence of any $\Z_2$ symmetry broken phase.
    }
The coinciding theta dependence of the vacuum energy,
given the relation (\ref {eq:map theta}) between $\theta$-angles,
also implies that the topological susceptibilities
of the two theories satisfy the correct relation for connected
correlation functions under large $N$ orbifold equivalence \cite {KUY2}.
For a $\Z_2$ projection, this is%
\footnote
    {
    For a $\Z_k$ projection connecting $U(kN)$ and $U(N)^k$ gauge theories,
    connected $L$-point correlators of neutral single-trace operators,
    normalized to have finite non-trivial large $N$ expectation values,
    are related by
    \begin {equation}
        \lim_{N\to\infty} (k N)^{2(L-1)} \,
	\langle {\cal O}_p(x_1) \cdots {\cal O}_p(x_L) \rangle_{\rm conn}^{(p)}
	=
        \lim_{N\to\infty} (k N^2)^{L-1} \,
	\langle {\cal O}_d(x_1) \cdots {\cal O}_d(x_L) \rangle_{\rm conn}^{(d)}
	\,,
    \label {eq:corr mapping}
    \end {equation}
    where the projection takes ${\cal O}_p \to {\cal O}_d$
    \cite {KUY2}.
    For the topological susceptibility,
    $
        {\cal O}_p = {1\over kN}\, \tr \, F \widetilde F/(16\pi^2)
    $
    and
    $
        {\cal O}_d = {1\over kN}\, \sum_{j=1}^k
	\tr\, F^j \widetilde F^j/(16\pi^2)
    $.
    One way to understand the different factors on either side of
    relation (\ref {eq:corr mapping}) is by analogy with the usual
    $\hbar \to 0$ limit.
    $L$-point connected correlators are proportional to $\hbar^{L-1}$.
    In the large $N$ limit under discussion, the dimension of the
    gauge group (or the number of gluons) plays the role of $\hbar$,
    namely,
    $(kN)^2$ for the parent theory and $k N^2$ for the daughter.
    Therefore, when comparing connected correlators between the two
    theories, one must scale the parent correlators by a factor
    of $k^{L-1}$, relative to the daughter, in order to compensate
    for the differing effective values of $\hbar$.
    \label {fn:conn}
    }
\begin {eqnarray}
    && \lim_{N \to \infty} \>
    \int d^4x \> \left\langle
	\frac 1{16\pi^2} \, \tr \, F \widetilde F(x) \>
	\frac 1{16\pi^2} \, \tr \, F \widetilde F(0)
    \right\rangle^\parent_{\rm conn}
\nonumber
\\ &=&
    \lim_{N \to \infty} \> {1 \over 2}
    \int d^4x \>
	\left\langle \!\vphantom {1 \over 2} \right.
	\sum_{j=1}^2
	\frac 1{16\pi^2} \, \tr \, F^j \widetilde F^j(x) \>
	\sum_{l=1}^2
	\frac 1{16\pi^2} \, \tr \, F^{l} \widetilde F^{l}(0)
    \left. \!\vphantom {1 \over 2} \right\rangle^\daughter_{\rm conn} \,.
\end {eqnarray}
With the correct mapping between parent and daughter theories,
there is no sign of any failure of
large $N$ orbifold equivalence in this example of a
$\Z_2$ projection of ${\cal N}\,{=}\,1$ supersymmetric Yang-Mills
theory, in infinite volume.
The apparent mismatch found by the authors of
Refs.~\cite{Gorsky-Shifman,ASV}
was due to an inappropriate assumption of coinciding vacuum energies
and/or theta angles.

\subsection {Anomalies}\label{sec:anomalies}

Gauge and gravitational contributions to the chiral anomaly
provide an instructive test of the consistency of the mapping between
parent and daughter theories.
The chiral currents in the parent and daughter theories are
given by 
\begin {eqnarray}
    J^\mu_p &\equiv& \frac 1{g_p^2} \, \tr\, \bar\lambda \, \bar\sigma^\mu \lambda \,,
\label {eq:J_p}
\\
\noalign {\hbox {and}}
    J^\mu_d &\equiv&
    \frac 1{g_d^2} \,
    \tr (
      \bar\lambda^1 \, \bar\sigma^\mu \lambda^1 +
      \bar\lambda^2 \, \bar\sigma^\mu \lambda^2 ) \,.
\label {eq:J_d}
\end {eqnarray}
One may repackage the two bifundamental Weyl fermions of the daughter theory
into a single Dirac fermion
$
    \Psi = \Bigl( { \lambda^1 \atop \bar\lambda^2 } \Bigr)
$,
in which case the daughter chiral current acquires the familiar form
$
    J^\mu_d
    =
    \frac 1{g_d^2} \, \tr\, \bar\Psi \gamma^\mu \gamma^5 \Psi
$.
The currents (\ref {eq:J_p}) and (\ref {eq:J_d}) are normalized to give
fermions equal charge assignments in the two theories,
so that the corresponding chiral charges
$Q = \int d^3x \> J^0(x)$ generate the same phase rotation
when acting on the fermion fields in either theory.
Under the orbifold projection, the chiral current $J_p^\mu$ is {\em not}
mapped to $J_d^\mu$, but rather
\begin {equation}
    J_p^\mu \to 2 \, J_d^\mu \,.
\label {eq:map J}
\end {equation}
This is consistent with the definitions (\ref {eq:J_p}) and (\ref {eq:J_d}),
combined with the relation (\ref {eq:map coupling}) between gauge couplings.
It may also be seen as a consequence of the fact that
generators of chiral symmetry transformations
must obey the same relation under orbifold projection,
namely $Q_p \to 2 \, Q_d$, as do the Hamiltonians of the two theories,
which generate time translations, or all the rest of the generators of
the infinite dimensional group which creates large $N$ coherent states
\cite {KUY2}.

The divergence of the chiral current in the parent theory
has a gauge field contribution of
$
    \frac {2N}{8\pi^2} \> \tr\, F_{\mu\nu} \widetilde F^{\mu\nu}
$
and, in curved space, a gravitational contribution of
$
    -\frac{N^2}{96\pi^2} \,
    R_{\mu\nu\alpha\beta} \, \widetilde R^{\mu\nu\alpha\beta}
$
\cite {F&S}.
(This is $2N^2$ times the contribution of a single Dirac fermion.)
The projection takes
$ \tr F \widetilde F $ to
$ \tr (F^1 \widetilde F^1 + F^2 \widetilde F^2) $,
so applying the orbifold projection to both sides of the
chiral anomaly equation in the parent gives
\begin {eqnarray}
    \nabla_\mu \, J^\mu_p
    &=&
    -
    \frac{N^2}{96\pi^2} \,
    R_{\mu\nu\alpha\beta} \, \widetilde R^{\mu\nu\alpha\beta}
    +
    \frac {2N}{8\pi^2} \> \tr\, F_{\mu\nu} \widetilde F^{\mu\nu}
\label {eq:chiral anom p}
\\
    \downarrow\quad && \kern 0.6in \downarrow \kern 1.3in \downarrow
\nonumber
\\
    2\, \nabla_\mu \, J^\mu_d
    &=&
    -
    \frac{N^2}{96\pi^2} \,
    R_{\mu\nu\alpha\beta} \, \widetilde R^{\mu\nu\alpha\beta}
    +
    \frac {2N}{8\pi^2} \>
    \tr\, ( F^1 \widetilde F^1 + F^2 \widetilde F^2 ) \,.
\label {eq:chiral anom d}
\end {eqnarray}
Dividing both sides of Eq.~(\ref {eq:chiral anom d}) by two
gives the correct chiral anomaly equation for the daughter theory, namely
$
    \nabla_\mu \, J^\mu_d
    =
    -
    \frac{N^2}{192\pi^2} \,
    R_{\mu\nu\alpha\beta} \, \widetilde R^{\mu\nu\alpha\beta}
    +
    \frac {N}{8\pi^2} \>
    \tr\, ( F^1 \widetilde F^1 + F^2 \widetilde F^2 )
$.
The coefficient of $R\tilde R$ 
is half as large in the daughter as in the parent,
because there are half as many fermionic degrees of
freedom in the daughter theory.

One may compare the trace anomalies of the two theories
in a completely analogous fashion.
The stress-energy tensors are related by
\begin {equation}
    T^{\mu\nu}_p \to 2 \, T^{\mu\nu}_d \,,
\label {eq:map T}
\end {equation}
with the factor of two appearing for the same reason as for the chiral
currents or the Hamiltonian.
The parent theory trace anomaly (in flat space, to lowest order),
and its projection to the daughter, is
\begin {eqnarray}
    (T_p)^\mu_\mu
    &=&
    \frac {6N}{16\pi^2} \> \tr\, F_{\mu\nu} F^{\mu\nu}
\\
    \downarrow\quad && \kern 0.6in \downarrow
\nonumber
\\
    2\, (T_d)^\mu_\mu
    &=&
    \frac {6N}{16\pi^2} \> \tr \, (F^1 F^1 + F^2 F^2) \,.
\label {eq:trace anom d}
\end {eqnarray}
Dividing both sides of Eq.~(\ref {eq:trace anom d}) by two
gives the correct trace anomaly equation for the daughter theory.

Finally, instead of considering the operator relations
(\ref {eq:chiral anom p}) and (\ref {eq:chiral anom d}),
one may alternatively examine the consistency of the
gravitational contribution to the chiral anomaly by considering
the three-point correlation function 
$\partial_\mu \langle J^\mu \, T^{\alpha\beta} \, T^{\gamma\delta} \rangle$,
since a linearized metric perturbation couples,
in either theory,
to the stress-energy tensor via a term $h_{\mu\nu}\, T^{\mu\nu}$.
Using the appropriate mappings (\ref {eq:map J}) and (\ref {eq:map T})
for these operators, plus the correct relation between connected
correlators, as described in footnote \ref {fn:conn},
one sees that the $\langle JTT \rangle$ correlator in the parent maps to twice
the $\langle JTT \rangle$ correlator in the daughter ---
correctly reflecting the
fact that the coefficient of $R\widetilde R$ in the chiral anomaly
of the daughter theory is half as large as in the parent.

It should be emphasized that the consistency of the above relations
among anomalies does not constitute a test of the validity of
non-perturbative equivalence between parent and daughter theories,
since chiral and trace anomalies
depend only on short distance perturbative physics
(and not on the $\Z_2$ symmetry realization in the daughter theory).
The fact that anomaly relations of the parent theory are mapped
onto the correct anomaly relations in the daughter is a logical
consequence of the perturbative equivalence between planar diagrams
\cite{Bershadsky-Johansen}, and merely serves to illustrate how
the correct mappings of gauge couplings, operators, and connected
correlators all fit together in a consistent fashion.

\subsection {Theta dependence and vacuum structure}

In this example of a $\Z_2$ projection of ${\cal N}\,{=}\,1$ supersymmetric
Yang-Mills theory, the parent theory (with no gluino mass)
has $2N$ distinct degenerate vacua, while the daughter theory has only $N$.
This would appear to be {\em prima-facie} evidence against a
non-perturbative large $N$ equivalence.
However, large $N$ orbifold equivalence only
applies to physical quantities which can be extracted from the
leading large $N$ behavior of the free energy density,
or from connected correlators of neutral single-trace gauge invariant
operators
({\em i.e.}, invariant under
the discrete projection symmetry in the parent, and the discrete
theory space symmetry in the daughter) \cite {KUY1,KUY2}.
As noted in the introduction, this includes mass spectra
and scattering amplitudes of particles in the relevant
neutral symmetry channels, but this information does not
directly yield a count of the number of vacua.%
\footnote
    {
    At low, but non-zero, temperature,
    the number of degenerate zero-temperature vacua appears as an overall
    multiplicative factor in the partition function.
    Hence, the number of vacua makes an
    $O(\ln N)$ contribution to the free energy density
    ${\cal F} \equiv -(\ln Z) / (\beta {\cal V})$,
    and the mis-match in the number of vacua contributes $\ln 2$ to
    the difference in $\beta {\cal V F}$ between the two theories.
    However, the differing spectra of particles in the non-neutral
    symmetry channels {\em also} produces $O(1)$ differences in the free
    energy density.
    Large $N$ orbifold equivalence, for the free energy, only applies to
    the leading $O(N^2)$ contribution (which happens to be temperature
    independent in the low-temperature confining phase).
    }

Despite the difference in the number of vacua,
large $N$ orbifold equivalence {\em is} applicable to all vacua
of the parent theory,
not just some subset of ``corresponding'' vacua
(as suggested in Ref.~\cite {Strassler}).
The resolution of this apparent paradox lies in the rescaling
(\ref {eq:map theta}) between the parent and daughter theta parameters.
The parent theta parameter $\theta_p$ is an angle which is only
defined modulo $2\pi$.
Because of this, the relation (\ref {eq:map theta}) between
parent and daughter theta parameters is necessarily double-valued,
\begin {equation}
    \theta_d = \half \, \theta_p \quad {\it or}\quad
    \theta_d = \half \, \theta_p + \pi \,,
\end {equation}
so the mapping between parent and daughter theories
is really a one-to-two mapping.

In the massless theory,
the choice of the daughter theta angle, for a given parent theta angle,
depends on which vacua in the parent theory one chooses to examine.
At, for example, $\theta_p = 0$,
the $N$ parent vacua for which the phase of
$\langle \lambda\lambda \rangle$ is an even multiple of $\pi/N$
map to the $N$ vacua in the daughter theory with $\theta_d = 0$,
while the other $N$ parent vacua, where
$\langle \lambda\lambda \rangle$ has a phase that
is an odd multiple of $\pi/N$,
map to the $N$ vacua of the daughter theory with $\theta_d = \pi$.
This is illustrated in Fig.~\ref {fig:theta-mapping}.

\begin{FIGURE}[t]
{
  \parbox[c]{\textwidth}
  {
  \begin{center}
  \includegraphics[width=2.4in]{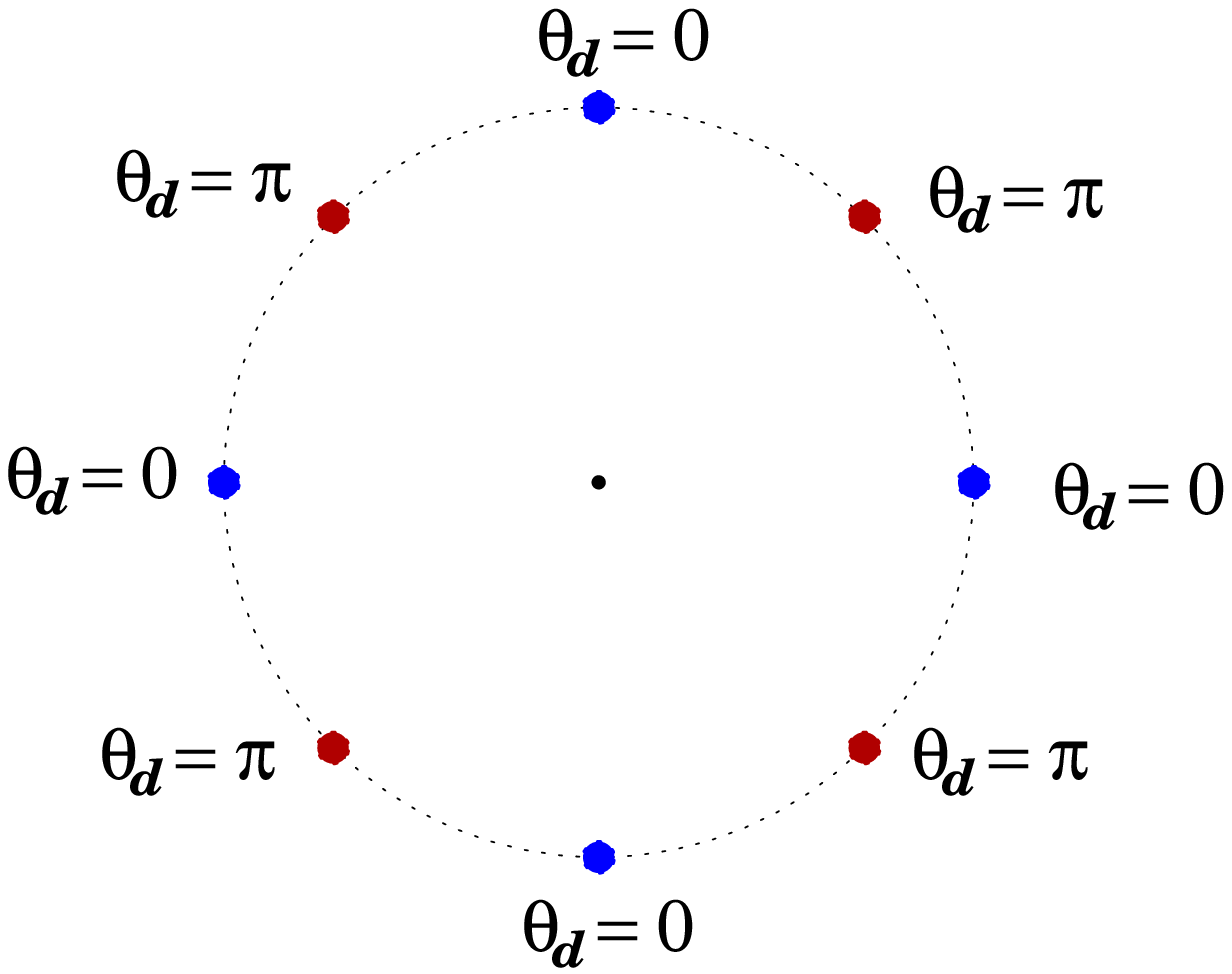}
  \caption
    {%
    The connection between the choice of vacua in the parent theory,
    plotted in the complex plane of $\langle \lambda\lambda\rangle^\parent$,
    and the choice of theta parameter for the associated daughter theory,
    illustrated for $N = 4$ and $\theta_p = 0$.
    }
  \end{center}
  }
\label{fig:theta-mapping}
}
\end{FIGURE}

\begin{FIGURE}[t]
{
  \parbox[c]{\textwidth}
  {
  \begin {center}
  \includegraphics[width=3.0in]{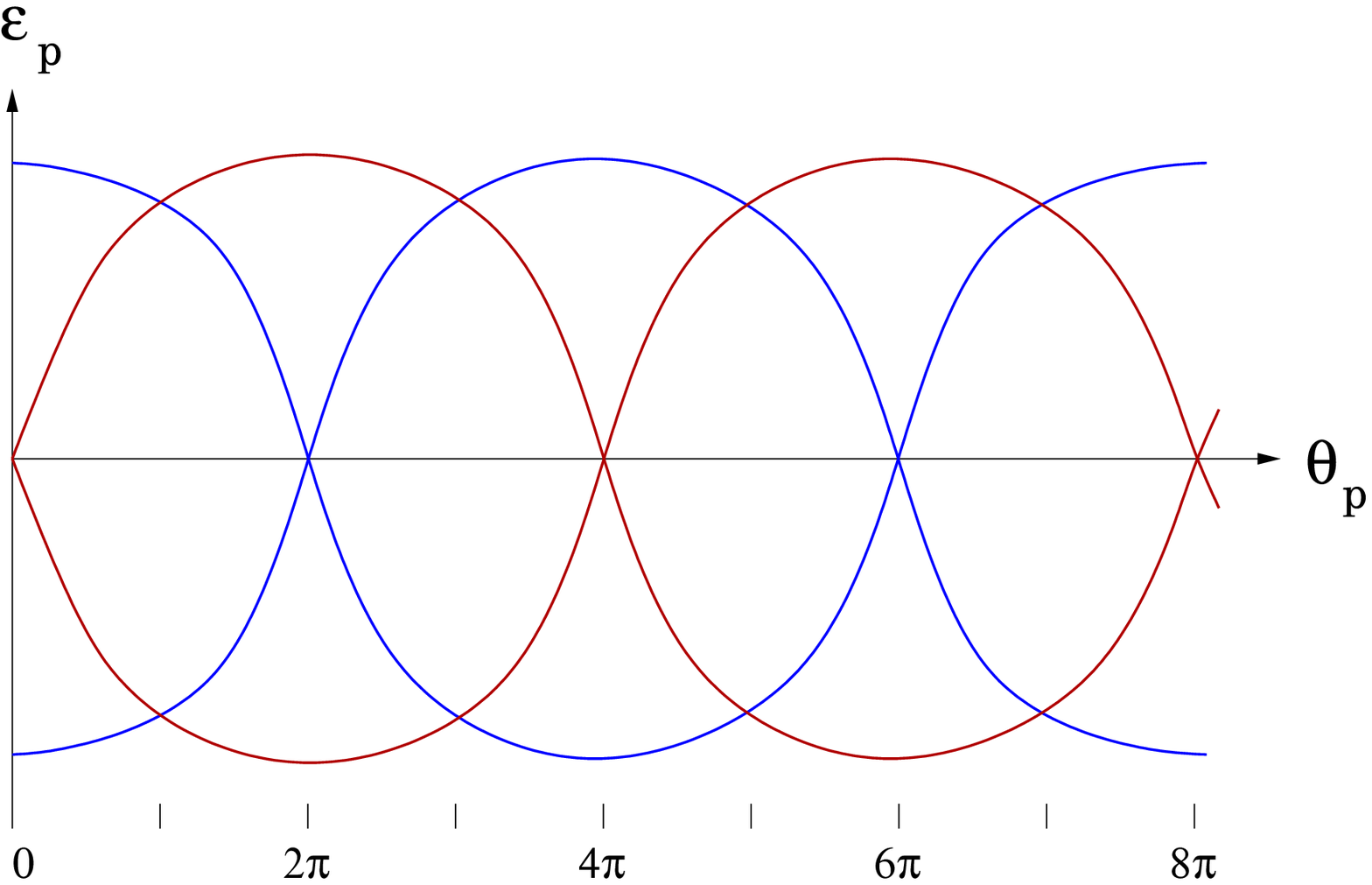}
  \hfil
  \includegraphics[width=3.0in]{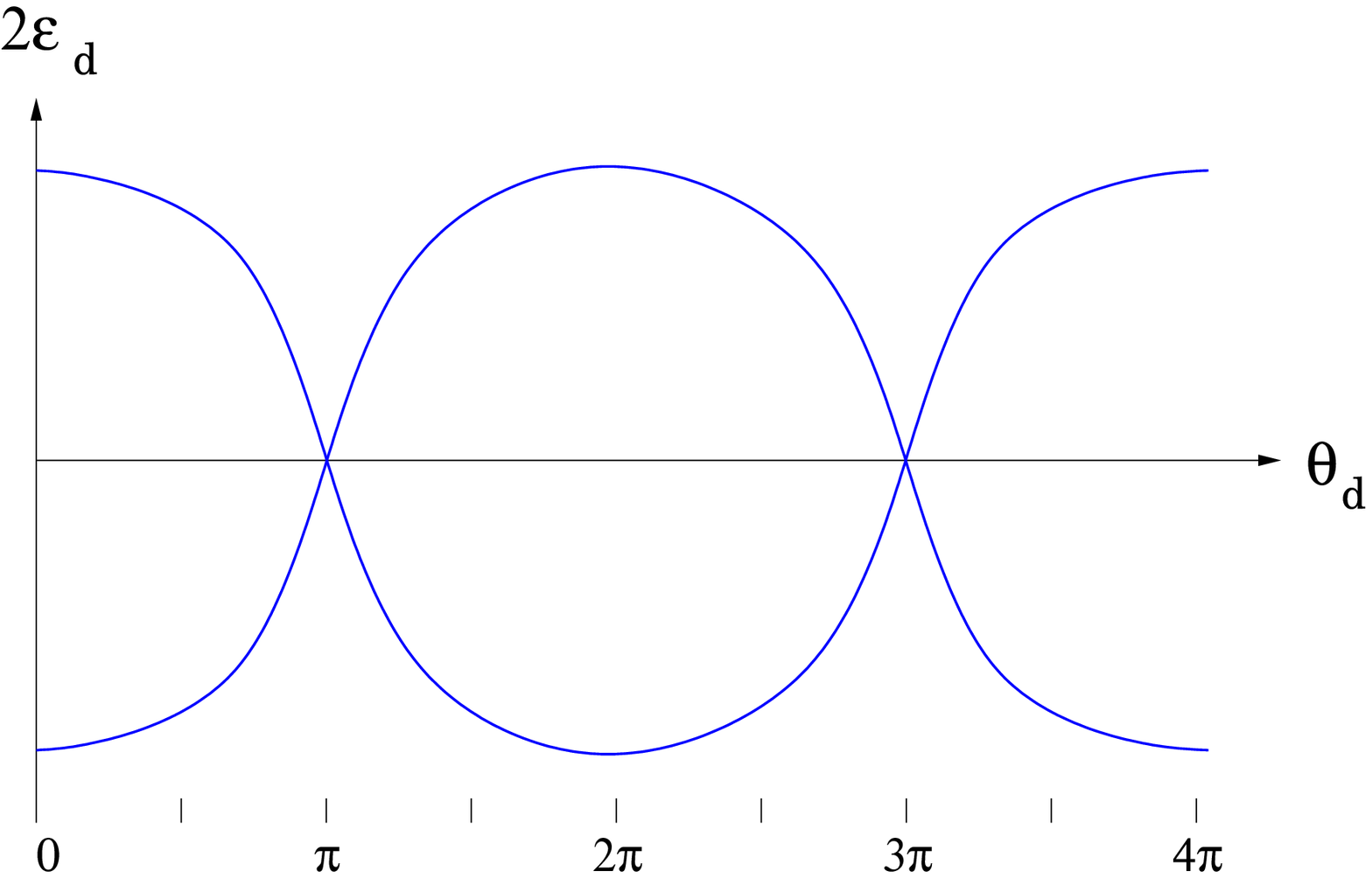}
  \caption
    {%
    Theta dependence of vacuum energy densities
    in the parent (left) and daughter (right) theories,
    illustrated for $N = 2$, when a small fermion mass is turned on.
    If the daughter theory graph is superimposed on the same
    graph shifted left (or right) by $\pi$, then the
    result coincides with the parent theory graph
    after rescaling the theta axis by a factor of two,
    illustrating the connection between the choice
    of vacua in the parent and the choice of $\theta$ in the daughter.
    }
  \end{center}
  }
\label{fig:theta-dependence}
}
\end{FIGURE}

With a small fermion mass turned on, the vacuum degeneracy of the massless
theories is lifted and the energy densities of the different vacua
acquire theta dependence as shown in
Eqs.~(\ref {eq:dE-parent}) and (\ref {eq:dE-daughter}),
and illustrated in Fig.~\ref{fig:theta-dependence}.
After rescaling $\theta_p$ by $1/2$,
half of the curves in the parent theory graph coincide with the
daughter theory curves, while the other half of the parent curves
coincide with the daughter theory curves shifted left (or right) by $\pi$.
The true ground state of the parent theory changes discontinuously
and the ground state energy density has a cusp when $\theta_p = \pi$
(mod $2\pi$).
At a given value of $\theta_p$, properties in the true ground state
of the parent theory map to corresponding properties in the ground
state of the daughter theory with $\theta_d$ equal to either
$\half \, \theta_p$, or $\pi + \half \, \theta_p$,
depending on which value yields the lower vacuum energy.
Physical quantities in both parent and daughter theories
have theta dependence which is periodic with period $2\pi$.
This would be inconsistent with the
$\theta_p = 2 \, \theta_d$ relation,
were it not for the two-to-one nature of the
mapping between daughter and parent theories.

\subsection {Domain walls}
\label{sec:walls}

Any theory which spontaneously breaks a discrete symmetry will have
stable domain walls which interpolate between different degenerate vacua.
In the $U(2N)$ supersymmetric parent theory, there are stable BPS domain
walls connecting any pair of vacua.
The tension of domain walls connecting vacua whose fermion condensate
phases differ by $\Delta k \, (\pi/N)$ is
\begin {equation}
    T_p(\Delta k) = C_p \, N^2 \, \sin \left({\pi \Delta k \over 2N}\right)
    \,, \qquad
    \Delta k = 1, \cdots, 2N{-}1 \,,
\end {equation}
where $C_p$ is a pure number times the strong scale $\Lambda^3$
\cite {Dvali}.
For large $N$, the tension of near-minimal domain walls
[for which $\Delta k$ is fixed as $N$ grows]
is $O(N)$,
while the tension of near-maximal domain walls
[with $\Delta k/N$ held fixed as $N$ grows]
is $O(N^2)$.
Domain walls with $\Delta k > 1$ are stable.
The binding energy density
relative to $\Delta k$ separate minimal domain walls,
$T_p(\Delta k) - \Delta k \> T_p(1)$,
is $O(1/N)$ for near-minimal walls and vanishes as $N \to \infty$,
while the binding energy of near-maximal walls is $O(N^2)$.

The domain wall tension should be viewed as the change in the ground
state energy, per unit area, produced by changing from ordinary
periodic boundary conditions to
``twisted'' periodic boundary conditions for which fermion fields
at, say, $z=-\infty$ are identified with the fields at $z=+\infty$ only
after applying a discrete chiral rotation which changes the phase
of $\tr\,\lambda\lambda$ by a chosen multiple of $\pi/N$.
Therefore, the appropriate mapping of domain wall tension between
parent and daughter theories is the same as for ground state energy
densities.
For our $\Z_2$ projection, this means the tension of a domain wall
in the parent theory should, for large $N$,
map to twice the tension of a corresponding
domain wall in the daughter theory,
\begin {equation}
    T_p = 2 \, T_d \times [1 + O(1/N^2)] \,.
\label {eq:Tp Td}
\end {equation}

There is an obvious puzzle with this correspondence:
the daughter theory (for a given value of $\theta_d$) has half as many
vacua as the parent, and consequently has fewer domain walls.
Choose, for convenience, $\theta_p = 0$.
Even-even domain walls --- which connect two vacua
whose condensate phases are both even multiples of $\pi/N$ ---
have an obvious mapping to domain walls in the daughter theory
with $\theta_d = 0$.
Similarly, odd-odd domain walls are mapped to domain walls in the daughter
theory with $\theta_d = \pi$.
If large $N$ equivalence holds for these domain walls,
then the large $N$ behavior of the tension $T_d(\Delta l)$
of domain walls in the daughter theory [connecting vacua whose
condensate phases differ by $\Delta l \, (2\pi/N)$]
must be
\begin {equation}
    T_d(\Delta l) = C_d \, N^2 \, \sin \left({\pi \Delta l \over N}\right)
    \times [1 + O(1/N^2)] \,,
    \qquad
    \Delta l = 1, \cdots, N-1 \,,
\label {eq:Td}
\end {equation}
with $C_d = \half \, C_p$,
so that
$
    T_p(2\Delta l) = 2 \, T_d(\Delta l)
$
as $N \to \infty$.

But what about even-odd domain walls in the parent theory
({\em i.e.}, walls for which $\Delta k$ is odd)?
These domain walls have no analogs in the daughter theories
with either value of $\theta_d$.
Does a lack of corresponding domain walls in the daughter theory
imply a failure of large $N$ equivalence between the parent
and daughter theories (which can only occur if the $\Z_2$
theory space symmetry is spontaneously broken in the daughter)?

No.
The resolution of this apparent problem lies in the divergence
of the domain wall tension as $N \to \infty$.
At finite $N$, any domain wall
will have fluctuations (quantum and thermal) in its position.
But the mean square size of such fluctuations is inversely
proportional to the domain wall tension.
The diverging tension in the large $N$ limit means that
domain walls, in this limit, are effectively {\em non-dynamical}.

\begin{FIGURE}[t]
{
  \parbox[c]{\textwidth}
  {
  \begin {center}
  \includegraphics[width=2.0in]{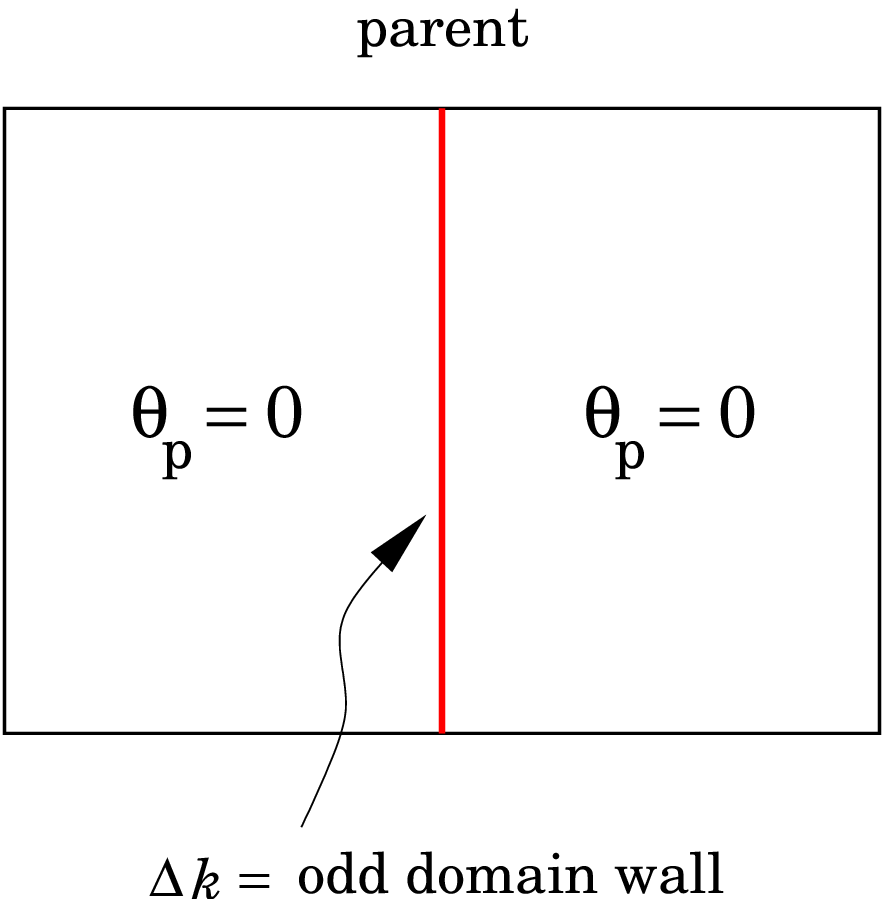}
  \hfil
  \includegraphics[width=2.0in]{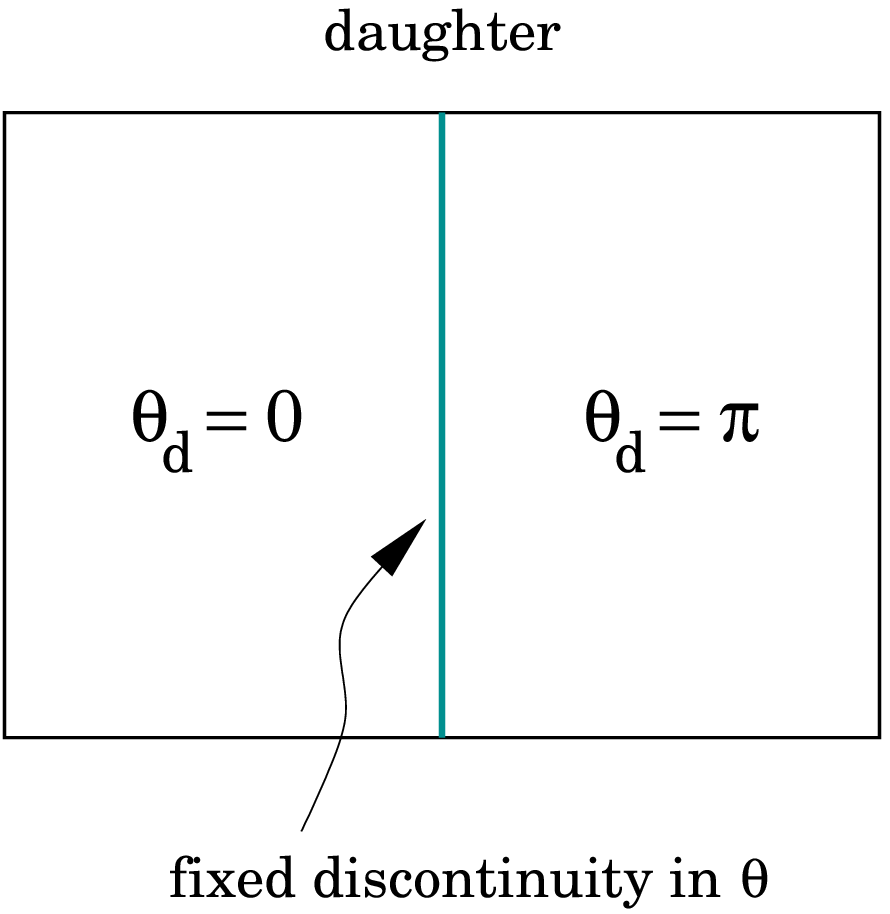}
  \caption
    {%
    Mapping of parent theory domain walls with $\Delta k$ odd
    onto a daughter theory whose theta parameter has a discontinuity
    of $\pi$ across a flat interface.
    }
  \end{center}
  }
\label{fig:odd walls}
}
\end{FIGURE}

Consider, for simplicity, the minimal domain wall in the parent theory,
connecting the $k=0$ and $k=1$ vacuum states (so the phase of the
condensate varies from 0 to $\pi/N$ as one crosses the wall),
with $\theta_p = 0$.
We have already argued that properties in the $k=0$ vacuum state
of the parent theory
map to properties in the corresponding vacuum state in the
daughter theory with $\theta_d = 0$,
while properties in the $k=1$ parent vacuum
map to properties in its corresponding vacuum state in the
daughter theory with $\theta_d = \pi$.
Therefore, it seems clear that properties of a non-dynamical
(as $N \to \infty$) domain wall interpolating between the $k=0$ and $k=1$
parent vacua should map onto a daughter theory in which
$\theta_d = 0$ on one side of the non-dynamical wall,
and $\theta_d = \pi$ on the other side,
as illustrated in Fig.~\ref {fig:odd walls}.
In other words, the multivaluedness of the mapping of $\theta$ parameters
between parent and daughter theories can become a discontinuous
non-translationally invariant mapping $\theta_p \to \theta_d(x)$
when considering domain walls, which are non-translationally invariant
static equilibrium states.%
\footnote
    {
    In a box with twisted periodic boundary conditions,
    as described above,
    the divergence of the domain wall energy
    as $N \to \infty$ implies that translation invariance
    is spontaneously broken in this limit,
    reflecting the continuous degeneracy associated with
    translating the domain wall in the normal direction
    At finite $N$ (and in a box with finite transverse size),
    this overall translation mode of a domain wall would be quantized,
    producing a zero momentum ground state and excited
    eigenstates with non-zero momentum $P_z$
    (for a domain wall parallel to the $x$-$y$ plane),
    and excitation energies equal to $P_z^2/(2M)$ where
    $M$ is the domain wall tension times its area.
    When the tension diverges as $N \to \infty$,
    this band of eigenstates becomes an infinitely degenerate ground state,
    signaling the spontaneous breaking of translation symmetry.
    }

In summary, there is a natural mapping of every domain wall in the
parent theory to a corresponding wall in an appropriate daughter theory,
provided one considers daughter theories where
$\theta_d = \half \, \theta_p$,
$\theta_d = \half \, \theta_p + \pi$,
{\em or}
$\,\theta_d = \half \, \theta_p + \pi \, n(x)$,
where $n(x)$ is a unit step function across a flat interface.
Since the daughter theory is not supersymmetric, no
exact evaluation of the resulting domain wall tensions is available.
But there is no apparent inconsistency with the prediction
of large $N$ orbifold equivalence, which requires
that relation (\ref {eq:Tp Td}) be satisfied for
all pairs of corresponding domain walls.

\section {\boldmath $\Z_k$ projections with $k > 2$}

One may contemplate starting with $U(kN)$ supersymmetric Yang-Mills
theory and applying a $\Z_k$ projection with $k > 2$.
Such projections were discussed in
Refs.~\cite{Schmaltz,Strassler,Gorsky-Shifman,ASV}.
If one chooses a projection group which is generated
by the $\Z_k$ gauge transformation
$
    \gamma \equiv
    {\rm diag}(\omega^0, \omega^1, \omega^2, \cdots, \omega^{k-1})
$,
with $\omega \equiv e^{2\pi i/k}$ times an $N \times N$ identity matrix,
then the daughter theory consists of $k$ decoupled copies of $U(N)$
supersymmetric Yang-Mills theory.
If one instead chooses a projection group which is generated
by the product of the gauge transformation $\gamma$ times
a discrete chiral rotation of the fermion $\lambda$ by $e^{2\pi i/k}$,
then the result is a $U(N)^k$ gauge theory with bifundamental Weyl
fermions transforming under ``nearest neighbor'' gauge groups,
whose theory space graph is illustrated in Fig.~\ref{fig:Zk theory space}.

\begin{FIGURE}[b]
{
  \parbox[c]{\textwidth}
  {
  \begin{center}
  \includegraphics[width=1.2in]{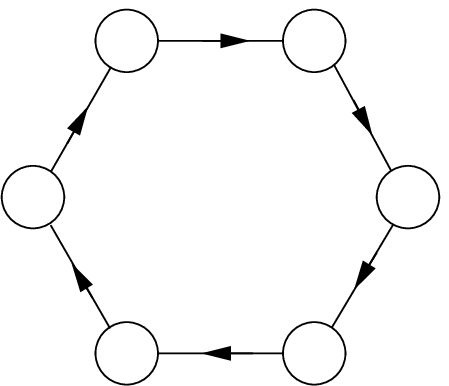}
  \caption
    {%
    The theory space for a $\Z_k$ projection of ${\cal N}\,{=}\,1$
    $U(kN)$ supersymmetric Yang-Mills theory yielding
    a non-supersymmetric $U(N)^k$ daughter theory with bifundamental fermions,
    for $k = 6$.
    Nodes represent individual $U(N)$ gauge group factors,
    while the arrows represent bifundamental fermions transforming
    in the fundamental and antifundamental representation, respectively,
    of the gauge group factors at either end.
    }
  \end{center}
  }
\label{fig:Zk theory space}
}
\end{FIGURE}

The daughter theory in this latter case is a chiral gauge theory
(when $k > 2$);
one cannot add gauge invariant mass terms for the fermions.
But the parent supersymmetric Yang-Mills theory is a non-chiral
vectorlike theory.
It would be truly remarkable if a large $N$ equivalence related
physical properties of these two theories.
Alas (but not surprisingly), this is not the case.
The discrete chiral symmetry transformation used to define this
orbifold projection
is part of the non-anomalous $\Z_{2kN}$ chiral symmetry
of the parent theory.
This chiral symmetry is known to break spontaneously down to $\Z_2$.
Therefore, any projection which involves chiral symmetry transformations
other than the $\Z_2$ symmetry of $(-1)^F$ will violate the
condition of unbroken projection symmetry in the parent
necessary for large $N$ orbifold equivalence.

The first choice of projection, generating $k$ decoupled copies
of $U(N)$ supersymmetric Yang-Mills theory from a single $U(kN)$ theory,
is a completely valid, but perhaps unremarkable, large $N$ equivalence.
Since there are no interactions between the multiple copies of
$U(N)$ theories in the daughter, the permutation symmetry
among the different copies
cannot possibly be spontaneously broken.
Neither can the projection symmetry in the parent theory,
since gauge symmetries cannot be spontaneously broken \cite{Elitzur}.
All gauge invariant operators in the parent are neutral under
this projection.
The mapping relates parent and daughter theories with coinciding
values of 't Hooft couplings and theta parameters rescaled by $k$,
that is
$g_p^2 = \frac 1k \, g_d^2$ and $\theta_p = k \, \theta_d$.

This large $N$ equivalence relating expectation
values in the parent and daughter theories
(or multi-point connected correlators when
scaled by appropriate powers of $(kN)^2$ in the parent
and $kN^2$ in the daughter \cite {KUY2}) merely reflects the existence
of the large $N$ limit in the original super-Yang-Mills theory.
One noteworthy feature is the mapping of discrete vacua.
The parent has $kN$ vacua while the daughter has $N^k$, of which $N$
are invariant under theory space permutation symmetry.
The difference between the number of parent vacua and permutation
symmetric daughter vacua is once again compensated by the $k$-fold
multivaluedness of the theta parameter mapping,
$\theta_d = (\theta_p + 2\pi j)/k$, $j = 0,1,\cdots,k{-}1$.
As a result, observables
(such as $\langle \lambda\lambda \rangle$)
in any vacuum state of the parent map to
corresponding observables in one of the permutation symmetric ground states
of the daughter theory for one of the choices of $\theta_d$.

The above discussion of domain walls also applies directly
to this $\Z_k$ projection.
Domain walls in the parent theory which connect vacua separated by multiples
of $k$ map onto domain walls in the daughter theory with
a single value of theta, while domain walls connecting vacua which
are not separated by multiples of $k$ map to daughter theories in
which $\theta$ is discontinuous (by a multiple of $2\pi/k$)
across a flat interface.

\section {Multiple fermion flavors}

One may add additional adjoint
representation fermions to the $U(2N)$ supersymmetric Yang-Mills parent theory,
to give a non-supersymmetric $U(2N)$ gauge theory with $n_f > 1$
adjoint Weyl fermions.%
\footnote
    {
    The leading term of the beta function is proportional to
    $(\frac{11}{3}- \frac{2}{3} \, n_f) \, N$,
    so preserving asymptotic freedom, which our discussion assumes,
    restricts the number of flavors to at most five.
    }
This theory has a non-anomalous
$SU(n_f) \times \Z_{4Nn_f}$ flavor symmetry,%
\footnote
    {
    More precisely, one should mod out
    from the $SU(n_f) \times \Z_{4Nn_f}$ product a
    $\Z_{n_f}$ subgroup which is common to both factors.
    And the unbroken symmetry is $O(n_f)$ rather than $SO(n_f)$
    if $n_f$ is odd.
    In either case, the center of the unbroken flavor symmetry group is the
    single $\Z_{2}$ generated by $(-1)^F$,
    which is relevant for understanding the pattern of discrete
    chiral symmetry breaking.
    }
which is expected to break spontaneously down to $SO(n_f)$, due to
the formation of a bilinear fermion condensate, giving rise
to $2N$ disjoint connected components in the vacuum manifold,
each of which is the coset space $SU(n_f)/SO(n_f)$.
The different components in the vacuum manifold are distinguished by the phase
of the determinant of the fermion condensate, in complete analogy to the phase
of the condensate in ${\cal N}\,{=}\,1$ supersymmetric Yang-Mills theory.
The breaking of the continuous chiral symmetry results
in $\half n_f (n_f{+}1)-1$ Goldstone bosons.

The same $\Z_2$ projection discussed in section 2, applied
to this multi-flavor theory, yields a
$U(N) \times U(N)$ daughter gauge theory with $n_f$ flavors
of bifundamental Dirac fermions.
This daughter theory has a
$SU(n_f)_L \times SU(n_f)_R \times \Z_{2Nn_f}$
flavor symmetry%
\footnote
    {
    A double-counted
    $\Z_{n_f} \times \Z_{n_f} \times \Z_2$ factor should similarly
    be divided out of this product symmetry group.
    There is also an overall $U(1)_B$ symmetry, but in a
    $U(N)\times U(N)$ theory with bifundamental fermions
    this is just part of the global gauge symmetry.
    }
which is expected to break spontaneously down to
the diagonal $SU(n_f)\times \Z_2$
[where the $\Z_2$ is $(-1)^F$],
resulting in $N$ disjoint connected
components in the vacuum manifold,
which are again distinguished by the phase of the determinant of the fermionic
condensate. The breaking of the continuous chiral symmetry
leads in this case to $n_f^2\,{-}\,1$ Goldstone bosons.

This mis-match in the number of Goldstone bosons in the multi-flavor
parent and daughter theories was previously noted in
Ref.~\cite{Gorsky-Shifman},
where it was asserted to be evidence for failure of large $N$ orbifold
equivalence in these multi-flavor theories.
However, this is a mis-interpretation.

The orbifold equivalence only applies to neutral symmetry channels
(those which are invariant under the $\Z_2$ projection symmetry
in the parent, and the $\Z_2$ theory space symmetry interchanging
gauge group factors in the daughter).
In the parent, all Goldstone bosons are neutral,
but in the daughter one may easily show that
$\half n_f (n_f{-}1)$ Goldstone bosons
are odd under the theory space $\Z_2$ symmetry,
while $\half n_f(n_f{+}1)-1$ of them are even ---
correctly matching the number in the parent theory.

The parent and daughter theories under consideration have both discrete
and continuous symmetry breaking. The latter gives rise to
Goldstone bosons, whereas the former is responsible for the emergence of
multiple connected components in the vacuum manifold and hence domain walls
which interpolate between different components.
The number of distinct connected components is $2N$ in the parent theory
and $N$ in the daughter. This situation, in essence, is identical to the
case of ${\cal N}\,{=}\,1$ supersymmetric Yang-Mills theory
(which is just the special case of $n_f=1$) and its $\Z_2$ projection.
This difference in the number of connected components is,
once again, fully consistent with non-perturbative large $N$ equivalence
due to the double valuedness of the mapping from parent to daughter
theta angles, as discussed in section \ref{sec:SYM}.

\section
{\boldmath $\Z_2$ projection of ${\cal N} \,{=}\,4$ supersymmetric Yang-Mills}

Numerous authors 
\cite{Tseytlin:1999ii,Adams:2001jb,Klebanov:1999ch,Nekrasov:1999mn,
Klebanov:1999um,Dymarsky:2005uh}
have examined the relation between
$U(2N)$ ${\cal N} \,{=}\,4$ supersymmetric Yang-Mills theory
and its $\Z_2$ projection which yields
a non-supersymmetric $U(N)^2$ theory with bifundamental fermions
and adjoint scalars, in both the weak-coupling limit using perturbative
methods, and in the strong-coupling limit using AdS/CFT duality.
It is natural to ask what lessons, if any, can be extracted from
the ${\cal N} \,{=}\,4$ case
that are relevant to the ${\cal N} \,{=}\,1$ case
discussed in this paper.

The validity of large $N$ equivalence
between parent and daughter theories reduces, in either case,
to the question of the realization of the $\Z_2$ symmetry
interchanging gauge group factors in the non-supersymmetric daughter theory.
In the case of the ${\cal N} \,{=}\, 4$ parent,
there is compelling evidence that this $\Z_2$ symmetry {\em is}\/
spontaneously broken in the daughter theory.
At weak coupling, the one-loop effective potential
for the adjoint scalars has a Coleman-Weinberg form,
with non-trivial minima in which the twisted operator
$\tr\, [(\Phi^1_a)^2 {-} (\Phi^2_a)^2]$
has a non-zero expectation value \cite {Tseytlin:1999ii,Adams:2001jb}.
At strong coupling, a dual description of the daughter theory
in terms of type 0 string theory was conjectured
\cite{Klebanov:1999ch}.
This theory has a tachyon which couples to the twisted operator
$
    {\cal O} \equiv
    \half \tr\, [F^1F^1 - F^2F^2]
    + \tr [(D\Phi_a^1)^2 - (D\Phi_a^2)^2]
    + \cdots
$
\cite{Klebanov:1999um},
thus suggesting that tachyon condensation signals
spontaneous breaking of the $\Z_2$ symmetry.
If one ignores this instability
(which must lead to a deformation of the geometry),
one finds an unphysical complex anomalous dimension for the operator
$\cal O$ at strong coupling (or small curvature).
Hence, for this $\Z_2$ projection of ${\cal N} \,{=}\, 4$
super-Yang-Mills theory,
it seems clear that large $N$ equivalence between parent
and daughter theories does not hold
due to spontaneous $\Z_2$ symmetry breaking in the daughter theory.

However, none of the analysis used for the ${\cal N} \,{=}\, 4$ case is
applicable to the $\Z_2$ projection of ${\cal N} \,{=}\, 1$ super-Yang-Mills.
In the latter case, there are no (non-composite) scalar fields and symmetry
breaking cannot be studied using weak-coupling perturbative methods.
Both parent and daughter theories are asymptotically free,
and neither has a known gravitational dual in which a supergravity
approximation is valid.
As always, the presence or absence of symmetry breaking depends on
the detailed dynamics of a theory.
So failure of large $N$ equivalence in the case of a $\Z_2$
projection of ${\cal N} \,{=}\, 4$ super-Yang-Mills tells one nothing
about the validity of large~$N$ equivalence in the ${\cal N} \,{=}\, 1$
case.

\section {Conclusions}

$\Z_k$ projections of ${\cal N}\,{=}\,1$
supersymmetric $U(kN)$ Yang-Mills theory,
producing $U(N)^k$ daughter theories,
provide very instructive examples of how large $N$ equivalence
between theories related by orbifold projection can work, or fail to work.
With the proper mapping of observables between parent and daughter theories,
large $N$ orbifold equivalence must hold
unless the chosen $\Z_k$ symmetry used to define the projection
is spontaneously broken in the parent, or the $\Z_k$
symmetry permuting gauge group factors is spontaneously broken
in the daughter \cite{KUY2}.

For $\Z_k$ projections with $k > 2$ that yield chiral theories,
large $N$ equivalence does fail due to chiral symmetry breaking
in the parent (and the use of a spontaneously broken chiral
transformation in the chosen projection).
So in this case, knowledge about properties of the parent theory,
in the large $N$ limit, cannot be used to infer properties
of the large $N$ daughter theory.

For $\Z_2$ projections of ${\cal N} = 1$ super-Yang-Mills,
the projection symmetry cannot be spontaneously broken in the parent,
so large $N$ equivalence must be valid unless the $\Z_2$
symmetry exchanging gauge group factors is spontaneously broken
in the daughter theory.
Therefore, if any consistency test of large $N$ equivalence can be
shown to fail, this would immediately imply spontaneous symmetry
breaking of the $\Z_2$ theory space symmetry in the daughter.

Contrary to previous claims, a comparison of vacuum structure
reveals complete consistency with large $N$ orbifold equivalence
for the $\Z_2$ projection of super-Yang-Mills (on $R^4$), or its
multiflavor generalizations.
Properties in any vacuum state of the parent correctly map
to properties in a corresponding vacuum state of the daughter,
provided one recognizes the multi-valuedness in the mapping
of the theta parameter between parent and daughter theories,
and provided one uses the correct mappings of parameters,
observables, and correlation functions between the two theories.

The correspondence between domain walls in parent and daughter
theories is somewhat subtle.
``Even" domain walls in the parent naturally correspond with
domain walls in the daughter theory at a given value of theta,
but ``odd'' domain walls map to daughter theories in which the
theta parameter has a discontinuity across a fixed flat interface.
This can be consistent due to the fact that domain walls become
non-dynamical, with vanishingly small fluctuations in position,
as $N \to \infty$.

A simple but essential point, which seems not to have been adequately
appreciated in some previous literature,
is that large $N$ equivalence between the (neutral sectors of)
parent and daughter theories does not mean strict equality
between all corresponding physical quantities.
Rather, it means that there is a well-defined mapping connecting
physical quantities in the two theories.
For a $\Z_k$ projection, this mapping necessarily involves rescaling
by appropriate powers of $k$.
The required rescalings are not arbitrary or difficult to understand.
The situation is completely analogous to comparing results in
semi-classical quantum theories at two different values of $\hbar$:
for quantities which scale like $\hbar^n$,
such as $n{+}1$ point connected correlators,
one must obviously rescale by $(\hbar_1/\hbar_2)^n$ before comparing.

In summary, we find no sign of any inconsistency which would
force one to conclude that large $N$ equivalence cannot hold
between the parent ${\cal N} = 1$ supersymmetric Yang-Mills theory and its
non-supersymmetric $\Z_2$ projection.
Available evidence supports the expectation that the $\Z_2$ theory space
symmetry remains unbroken in the daughter theory (on $R^4$),
and hence that there is a non-perturbative large $N$ equivalence
between these theories.
Verifying, or disproving, this expectation via lattice simulations
would be desirable and should be feasible.

\subsubsection*{Note added:}

The recent preprint \cite {AGS} discusses the same $\Z_2$ projection
of super-Yang-Mills theory, and examines a number of physical
quantities which can be used as consistency tests of large $N$ equivalence
between parent and daughter theories.
The authors of Ref.~\cite {AGS} argue that all their tests fail,
and therefore assert that
spontaneous breaking of $\Z_2$ symmetry must occur in the daughter theory.
However, the apparent inconsistencies noted in Ref.~\cite {AGS}
are all consequences of incorrect mappings of observables and/or
connected correlators between parent and daughter theories.
When the correct mappings are used, these apparent inconsistencies
disappear.%
\footnote
  {
  The supposed mis-match in the vacuum structure at $\theta_p = 2\pi$,
  discussed in section 2 of Ref.~\cite {AGS},
  is a consequence of overlooking the double-valued nature of
  the mapping from $\theta_p$ to $\theta_d$.
  With a non-zero mass turned on,
  it is clear that properties in the unique ground state
  of the parent theory at $\theta_p = 2\pi$
  correspond to properties in the unique ground state
  of the daughter theory at $\theta_d = 0$, not at $\theta_d = \pi$.
  The assertion (on pg.~7) that ``the partition functions for
  parent and daughter must coincide at large $N$'' is also incorrect.
  Both partition functions diverge exponentially as $N \to \infty$.
  The correct statement of equivalence for the partition functions is
  that the ratio of their free energies must approach two (not one)
  as $N \to \infty$, or
  $
    \lim_{N\to\infty} (2N)^{-2} \ln Z_p
    =
    \lim_{N\to\infty} (2N^2)^{-1} \ln Z_d
  $.

  The claimed inconsistency in
  gravitational contributions to the axial anomaly,
  discussed in section 3 of Ref.~\cite {AGS},
  is a result of the use of incorrect mappings between
  axial currents, stress-energy tensors,
  and topological charge densities.
  As discussed in section \ref{sec:anomalies} of this paper,
  when the correct relations are used
  the chiral anomaly in the parent theory
  (including both gauge and gravitational contributions)
  is properly mapped to the chiral anomaly in the daughter theory,
  as it must be for purely perturbative equivalence to be valid.

  The discussion in section 4 of Ref.~\cite {AGS}
  of $\langle \tr(F^1 F^1 + F^2 F^2) \rangle$ as an order parameter
  for the massless theory whose non-zero value, if $O(N)$ [not $O(N^2)$],
  would signal failure of large $N$ equivalence and hence
  imply $\Z_2$ symmetry breaking in the daughter theory is fine ---
  except that there is no evidence that this expectation
  value, suitably renormalized, is non-vanishing at $O(N)$.

  The comparison of domain wall tensions in section 5
  asserts that $\tr \, FF$ maps to $\tr(F^1 F^1 + F^2 F^2)/\sqrt 2$,
  and then asserts that valid large $N$ equivalence requires
  that parent and daughter domain wall tensions coincide.
  Neither is correct;
  $\tr \, FF$ maps to $\tr(F^1 F^1 + F^2 F^2)$ with no $\sqrt 2$,
  and the tension in the parent theory should be compared
  with twice the tension of the corresponding wall in the daughter,
  as explained in section \ref{sec:walls} above.
  Once one corrects the mis-understanding regarding coinciding
  vacuum structure in parent and daughter theories, we fail to
  see any basis for concluding that domain walls in the
  daughter theory can split up into ``fractional domain walls.''
  The same mis-understanding regarding the appropriate mappings
  of $\tr(FF)$ and $\tr(F\tilde F)$, and of vacuum energies
  (namely ${\cal E}_\parent \to 2 \, {\cal E}_\daughter$)
  are responsible for the apparent inconsistencies presented
  in section 6 and the appendix of Ref.~\cite {AGS}.

  In summary, the assertions in Ref.~\cite{AGS} that
  ``Ample evidence ... establishes ... nonperturbative nonequivalence''
  and that
  ``this is the first example of spontaneous breaking of a discrete
  symmetry in a strongly coupled gauge theory ever established
  analytically in four dimensions'' are unfounded.
  Spontaneous breaking of the $\Z_2$ theory space symmetry of
  the daughter theory (in infinite volume) produced by a $\Z_2$
  projection of ${\cal N}\,{=}\,1$ super-Yang-Mills
  remains a logical possibility --- but everything known about
  this theory, so far, is consistent with the absence of such
  symmetry breaking, and hence with the validity of large $N$
  equivalence in this case.
  }

\vfill

\begin{acknowledgments}

This work was supported, in part,
by the U.S. Department of Energy under Grant Nos.~DE-FG02-96ER40956
and DE-FG02-91ER40676,
and the National Science Foundation under Grant No.~PHY99-07949.

\end{acknowledgments}



\begin {thebibliography}{99}

\bibitem{Bershadsky-Johansen}
    M.~Bershadsky and A.~Johansen,
    {\it Large N limit of orbifold field theories,}
    \npb{536}{1998}{141},
    \hepth{9803249}.

\bibitem{Schmaltz}
    M.~Schmaltz,
    {\it Duality of non-supersymmetric large N gauge theories,}
    \prd{59}{1999}{105018},
    \hepth{9805218}.

\bibitem{Strassler}
    M.~J.~Strassler,
    {\it On methods for extracting exact non-perturbative results in
    non-supersymmetric gauge theories,}
    \hepth{0104032}.

\bibitem{Dijkgraaf-Neitzke-Vafa}
   R.~Dijkgraaf, A.~Neitzke and C.~Vafa,
   {\it Large N strong coupling dynamics in
   non-supersymmetric orbifold field  theories,}
   \hepth{0211194}.

\bibitem{KUY1}
     P.~Kovtun, M.~\"Unsal and L.~G.~Yaffe,
     {\it Non-perturbative equivalences among large $N_c$ gauge theories
     with adjoint and bifundamental matter fields,}
     \jhep{0312}{2003}{034},
     {\tt hep-th/0311098}.

\bibitem{KUY2}
     P.~Kovtun, M.~\"Unsal and L.~G.~Yaffe,
     {\it Necessary and sufficient conditions for non-perturbative
     equivalences of large $\Nc$ orbifold gauge theories,}
     {\tt hep-th/0411177}.

\bibitem{Gorsky-Shifman}
   A.~Gorsky and M.~Shifman,
   {\it Testing nonperturbative orbifold conjecture,}
   \prd{67}{2003}{022003},
   \hepth{0208073}.

\bibitem{Tong}
   D.~Tong,
   {\it Comments on condensates in non-supersymmetric
   orbifold field theories,}
   \jhep{0303}{2003}{022},
   \hepth{0212235}.

\bibitem{ASV}
    A.~Armoni, M.~Shifman and G.~Veneziano,
    {\it From super-Yang-Mills theory to QCD: planar equivalence and its
    implications,}
    \hepth{0403071}.

\bibitem{Erlich-Naqvi}
   J.~Erlich and A.~Naqvi,
   {\it Nonperturbative tests of the parent/orbifold correspondence
   in  supersymmetric gauge theories,}
   \jhep{0212}{2002}{047},
   \hepth{9808026}.

\bibitem{Tseytlin:1999ii}
  A.~A.~Tseytlin and K.~Zarembo,
  {\it Effective potential in non-supersymmetric $SU(N) \times SU(N)$
   gauge theory  and interactions of type 0 D3-branes,}
  \plb {457}{1999}{77},
  \hepth{9902095}.

\bibitem{Adams:2001jb}
  A.~Adams and E.~Silverstein,
  {\it Closed string tachyons, AdS/CFT, and large N QCD,}
  \prd{64}{2001}{086001},
  \hepth{0103220}.

\bibitem{Klebanov:1999ch}
  I.~R.~Klebanov and A.~A.~Tseytlin,
  {\it A non-supersymmetric large N CFT from type 0 string theory,}
  \jhep {9903}{1999}{015},
  \hepth{9901101};

\bibitem{Nekrasov:1999mn}
  N.~Nekrasov and S.~L.~Shatashvili,
  {\it On non-supersymmetric CFT in four dimensions,}
  \prep{320}{1999}{127},
  \hepth{9902110}.

\bibitem{Klebanov:1999um}
  I.~R.~Klebanov,
  {\it Tachyon stabilization in the AdS/CFT correspondence,}
  \plb{466}{1999}{166},
  \hepth{9906220}.

\bibitem{Dymarsky:2005uh}
  A.~Dymarsky, I.~R.~Klebanov and R.~Roiban,
  {\it Perturbative search for fixed lines in large N gauge theories,}
  \hepth{0505099}.

\bibitem{Witten}
  E.~Witten,
  {\it Branes and the dynamics of {QCD},}
  \npb{507}{1997}{658},
  \hepth{9706109}.

\bibitem{Elitzur}
  S.~Elitzur,
  {\it Impossibility of spontaneously breaking local symmetries,}
  \prd{12}{1975}{3978},

\bibitem{F&S}
  See, for example,
  K.~Fujikawa and H.~Suzuki,
  {\it Path integrals and quantum anomalies,}
  Oxford, UK: Clarendon (2004).

\bibitem{Dvali}
  G.~R.~Dvali and M.~A.~Shifman,
  {\it Domain walls in strongly coupled theories,}
  \plb{396}{1997}{64};
  [Erratum-ibid.\ {\bf B 407} (1997) 452],
  \hepth{9612128}.

\bibitem{AGS}
  A.~Armoni, A.~Gorsky and M.~Shifman,
  {\it Spontaneous Z(2) symmetry breaking in the orbifold daughter of
  ${\cal N} = 1$
  super-Yang-Mills theory, fractional domain walls and vacuum structure,}
  \hepth{0505022}.

\end {thebibliography}
\end {document}